\documentclass[sn-aps]{sn-jnl}

\usepackage{caption}
\usepackage{multirow}
\usepackage{float}
\usepackage[sectionbib]{bibunits}
 
\jyear{2023}%

\theoremstyle{thmstyleone}%
\theoremstyle{thmstyletwo}%

\theoremstyle{thmstylethree}%

\begin{document}

\title[LaFeO$_3$/SrTiO$_3$ superlattice ]{
Observation of Mermin-Wagner behavior in LaFeO$_3$/SrTiO$_3$ superlattices
}


\author*[1]{\fnm{Michal} \sur{Kiaba}}\email{kiaba@mail.muni.cz}

\author[2]{\fnm{Andreas} \sur{Suter}}
\author[2]{\fnm{Zaher} \sur{Salman}}
\author[2]{\fnm{Thomas} \sur{Prokscha}}
\author[3]{\fnm{Binbin} \sur{Chen}}
\author[4]{\fnm{Gertjan} \sur{Koster}}

 \author[1,5]{\fnm{Adam} \sur{Dubroka}}


\affil*[1]{\orgdiv{Department of Condensed Matter Physics}, \orgname{Masaryk University}, \orgaddress{\street{Kotlářská 2}, \city{Brno}, \postcode{61137},  \country{Czech republic}}}

\affil[2]{\orgdiv{Laboratory for Muon-Spin Spectroscopy}, \orgname{Paul Scherrer Institute}, \city{Villigen}, \country{Switzerland}}

\affil[3]{Key Laboratory of Polar Materials and Devices (MOE) and Department of Electronics, East China Normal University, Shanghai 200241, China}

\affil[4]{MESA+ Institute for Nanotechnology, University of Twente, 7500 AE Enschede, The 
Netherlands}

\affil[5]{Central European Institute of Technology, Brno University of Technology, 612 00 Brno, Czech Republic}

\def\lfo{LaFeO\ensuremath{_3}}
\def\sto{SrTiO\ensuremath{_3}}
\def\tn{\ensuremath{T_{\rm N}}}
\def\muSR{$\mu$SR}


\begin{bibunit}
 \abstract{
 
Two-dimensional magnetic materials  attract a lot of attention since they potentially exhibit new magnetic properties due to, e.g., strongly enhanced spin fluctuations. However, the suppression of the  long-range magnetic order in two dimensions due to long-wavelength spin fluctuations, as suggested by the Mermin-Wagner theorem,  has been questioned for finite-size laboratory samples.  Here we study the magnetic properties of a dimensional crossover in superlattices composed of the antiferromagnetic LaFeO$_3$ and SrTiO$_3$  that, thanks to their large lateral size, allowed examination using a sensitive magnetic probe --- muon spin rotation spectroscopy. We show that the iron electronic moments in superlattices with 3 and 2 monolayers of \lfo\ exhibit a static  antiferromagnetic order. In contrast, in the superlattices with single \lfo\ monolayer, the moments do not order and fluctuate to the lowest measured temperature as expected from the Mermin-Wagner theorem. Our work shows how dimensionality can be  used to tune the magnetic properties of ultrathin films.

 }
%
%
%




\maketitle

The properties of magnetic films with thickness in the nanoscale have been a long-standing research topic. The theory of critical behavior predicts that the phase transition temperature should decrease with decreasing film thickness~\cite{Fisher1972}, which was observed in several cases \cite{Schneider1990, Fullerton1995,  Ambrose1996, Abarra1996, Zhang2001, Lang2006}. 
In the 2-dimensional (2D) limit, Mermin and Wagner~\cite{Mermin1966} extended the initial idea of Hohenberg~\cite{Hohenberg1967} for a superconductor and predicted 
complete suppression of the long-range magnetic order in models with continuous rotational symmetries (i.e., with the Heisenberg or XY spin Hamiltonian) at finite temperature due to long-wavelength fluctuations. Importantly, this prediction is strictly valid only for the thermodynamic limit, i.e., for samples with laterally infinite sizes. However, since the divergence of the fluctuations in 2D case is only slow (logarithmic in sample size), it was  suggested that for any finite-size laboratory samples, the phase order is preserved for superconductivity~\cite{Palle2021} and even for magnetism~\cite{Jenkins2022}.

The discovery of magnetic van der Waals materials allowed the investigation of magnetism in samples with thickness down to a single monolayer \cite{Burch2018}. 
For example, it was reported that in samples of bulk antiferromagnet NiPS$_3$ that are two or more monolayers thick, the magnetic order is preserved, whereas it is suppressed in a single monolayer sample~\cite{Kim2019}. Since the Hamiltonian of NiPS$_3$ has the XY symmetry, this behavior thus follows the prediction of the Mermin-Wagner theorem rather than the suggestions for preserving the long-range order~\cite{Jenkins2022}. However, due to the small lateral size  of the  single monolayer NiPS$_3$ samples obtained by the exfoliation, the antiferromagnetic order was probed relatively indirectly by Raman spectroscopy via coupling of a phonon to a magnon mode~\cite{Kim2019}.

To test the Mermin-Wagner behavior using a magnetic probe, we study the magnetic properties of three to two-dimensional crossover in superlattices composed of antiferromagnetic LaFeO$_3$ separated by  nonmagnetic SrTiO$_3$ layers.  Bulk \lfo\ is a prototypical perovskite antiferromagnetic insulator with Heisenberg symmetry of the spin Hamiltonian~\cite{McQueeney2008} and with the highest Neel temperature (\tn) of 740 K among {\it Re}FeO$_3$  materials~\cite{Eibschutz1967}, where \textit{Re} stands for rear earth. 
It has a high magnetic moment of almost 5 $\mu_{\rm B}$ per Fe$^{3+}$ ion and the G-type structure of the antiferromagnetic state (where each spin is aligned opposite to the nearest neighbor),   thus the antiferromagnetic order is expected to be relatively robust. Thanks to the advancement in deposition  technology, it is possible to fabricate heterostructures  with sharp interfaces that are composed of  perovskite oxides with various order parameters,  including magnetism, ferroelectricity, and superconductivity \cite{Hwang2012,Chen2017}. Perovskite oxide heterostructures are also promising for applications since they can be used in large-scale samples and devices (see e.g. Refs.~\cite{Ueno2008, Bell2009, Kozuka2011, Rong2018}).  Using  pulsed laser deposition, we  fabricated superlattices with  1, 2, and 3 monolayers of  LaFeO$_3$   separated by a non-magnetic spacer of 5 monolayers of SrTiO$_3$ with a large lateral size of  $10\times10$~mm$^2$ that allowed their investigation using a sensitive magnetic probe --- low-energy muon spin rotation spectroscopy~\cite{Prokscha2008}. 


\begin{figure}[!pt]
    \centering
    \includegraphics[width=11.5cm]{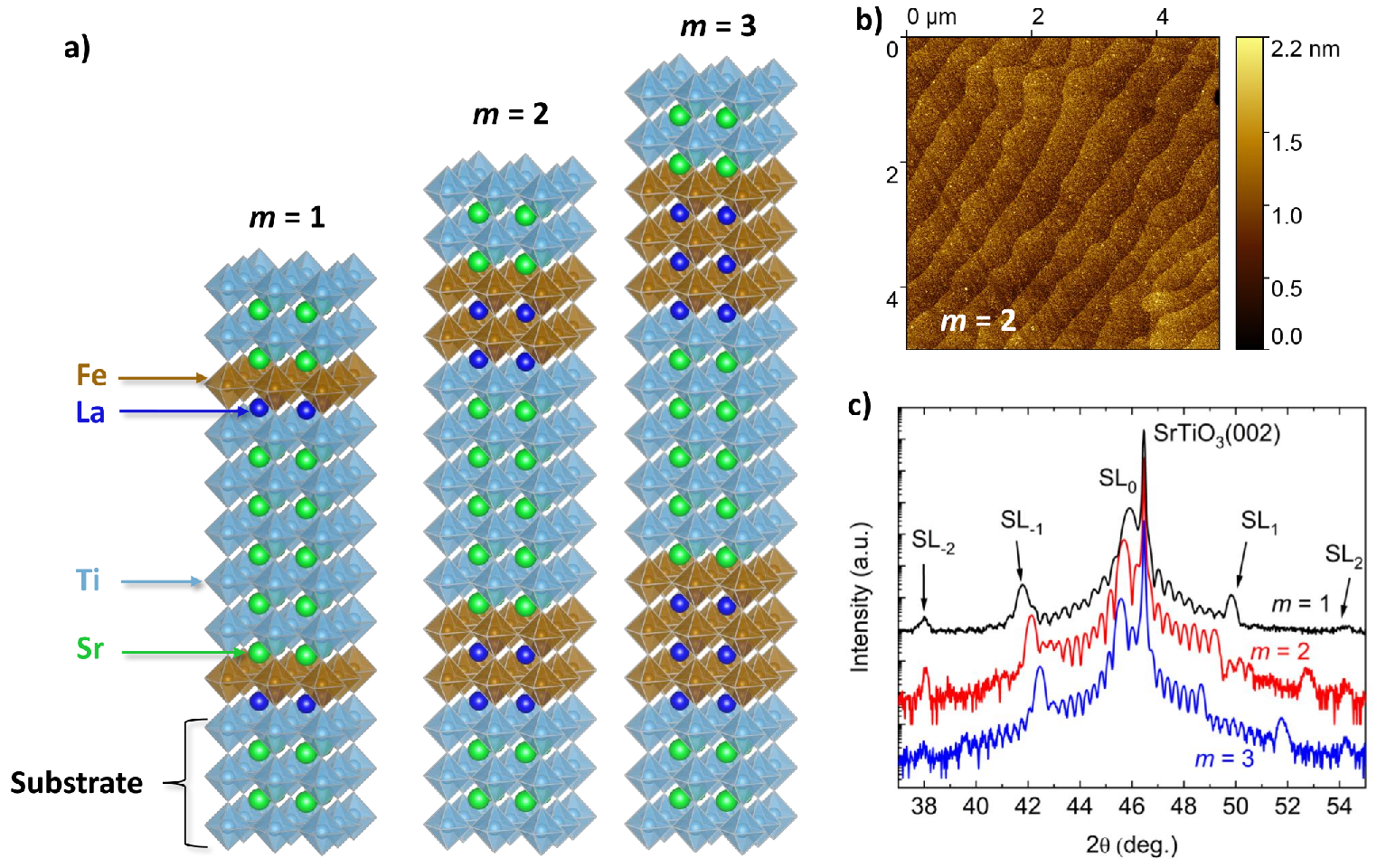}
    \caption{\textbf{Structural characterization of the superlattices.}
    a) Scheme of [(LaFeO$_3$)$_m$/(SrTiO$_3$)$_5$]$_{10}$ superlattices near the surface of TiO-terminated SrTiO$_3$ (001) substrate. b) The surface morphology of the $m=2$ superlattice determined by an atomic force microscope. c)~\mbox{X-ray} diffraction spectra  exhibiting
    zero (SL$_0$), first (SL$_1$,  SL$_{-1}$) and the second  diffraction peaks  (SL$_2$,  SL$_{-2}$) due to the (LaFeO$_3$)$_m$/(SrTiO$_3$)$_5$ bilayer occurring near the (002) diffraction  of the {SrTiO$_3$} substrate.
}
    \label{fig:Figure1}
\end{figure}

To enhance the signal in the muon spin rotation experiment, we prepared  superlattices denoted as [(LaFeO$_3$)$_m$/(SrTiO$_3$)$_5$]$_{10}$, where a bilayer with $m=1,2$ or 3 monolayers of LaFeO$_3$ and five monolayers of SrTiO$_3$ is repeated  10 times. The scheme of the ideal superlattice structure near the interface with the TiO-terminated SrTiO$_3$ (001) substrate is shown in Fig.~\ref{fig:Figure1}(a). Figure~\ref{fig:Figure1}(b) displays the surface morphology of the $m=2$ superlattice measured by an atomic force microscope, which exhibits a  flat surface with single unit cell steps copying those of the substrate. The X-ray diffraction spectra, see Fig.~\ref{fig:Figure1}(c), exhibits  zero (SL$_0$), first (SL$_1$,  SL$_{-1}$) and the second  superlattice diffraction peaks  (SL$_2$,  SL$_{-2}$) due to the (LaFeO$_3$)$_m$/(SrTiO$_3$)$_5$ bilayer, which depict the high structural quality of the superlattices. The  thickness of (LaFeO$_3$)$_m$/(SrTiO$_3$)$_5$ bilayer determined from the first order diffraction  peak follows very well the estimates based on the lattice constant of SrTiO$_3$ and LaFeO$_3$, see  Supplementary Fig.~\ref{fig:Figure9}(a).

Investigations of magnetic properties of  ultrathin antiferromagnetic layers is a challenging task because of their zero (or very small) average magnetic moment compared to the large total  diamagnetic moment of the substrate. To probe the magnetic properties of our superlattices, we have used muon spin rotation spectroscopy, which is sensitive to even very weak local magnetic fields and can distinguish between static and dynamic behavior. We performed the experiments with a low-energy (2~keV) muon beam~\cite{Prokscha2008,Suter2023}, where spin-polarized muons are implanted into the sample only within about 25 nm deep from the surface, see Supplementary Fig.~\ref{fig:Figure9}(b).
Any  magnetic field component transverse to the muon spin direction causes its precession  with the Lamour frequency  $\omega_L=\gamma_{\mu}B$, where $\gamma_\mu=ge/2m_\mu$ is the gyromagnetic ratio of the muon and $B$ is the magnitude of the local magnetic field. The time dependence  of polarization of the muon spin ensemble (the so-called asymmetry) is measured  thanks to the muon decay into  a positron  preferentially emitted along the muon spin~\cite{Blundell1999}.

\begin{figure}[t]
    \centering
    \includegraphics[width=11cm]{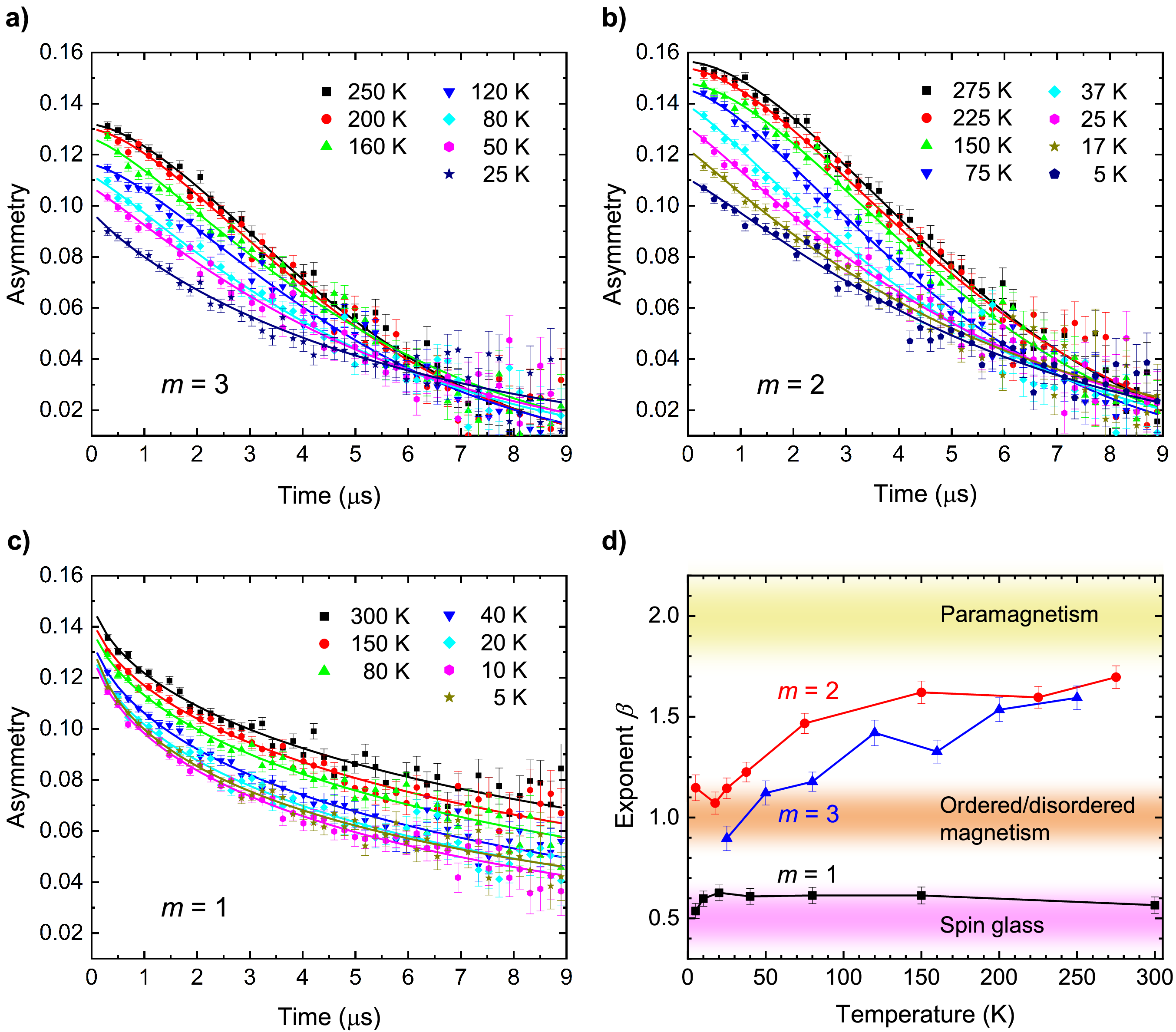}
    \caption{\textbf{Zero field muon spin rotation.}
    Time evolution of the zero-field muon spin polarisation  of [(LaFeO$_3$)$_m$/(SrTiO$_3$)$_5$]$_{10}$ superlattices with (a) $m$ = 3, 
    (b) $m$ = 2, and (c) $m$ = 1. Error bars represent one standard deviation, and solid lines represent fit by the stretched exponential function~(\ref{eq:1}). (d) Exponent $\beta$ of the stretched exponential as a function of temperature. Lines represent a guide to the eye. Colored regions show expected values of $\beta$ for different types of magnetic states.}
    \label{fig:Figure2}
\end{figure}

\subsection*{Zero field muon spin rotation}

Figure \ref{fig:Figure2} shows results from the muon spin rotation experiment in zero magnetic field.  The time dependence of the muon spin polarization of the superlattices with $m=$ 3 and 2, see Figs.~\ref{fig:Figure2}(a) and  \ref{fig:Figure2}(b), respectively, exhibit at high temperature a concave Gaussian-like profile and a transition to a faster exponential-like relaxation  at lower temperatures. This behavior is consistent with the following qualitative picture: at high temperatures, \lfo\ layers are in a paramagnetic state where the iron electronic moments are fluctuating too fast to be followed by muons, and thus the depolarization is mainly due to the  nuclear moments~\cite{Hayano1979}. With decreasing temperature, the iron electronic moments, that are much larger than the nuclear moments, start  ordering, which manifests as a drop of the initial asymmetry and a faster relaxation. In contrast, the asymmetry of the $m= 1$ superlattice  shown in Fig.~\ref{fig:Figure2}(c) is qualitatively different because even at high temperatures, it exhibits a faster depolarization with a convex profile.  Such behavior indicates that even at high temperatures, the iron electronic moments fluctuate relatively slowly, which masks the fields due to the nuclear moments.

To get a more quantitative insight, we analyzed the zero field asymmetry, $A_{\rm {ZF}}(t)$,  with the phenomenological stretched exponential function~\cite{Fowlie2022, Cantarino2019, Mustonen2018, Bert2006, Keren2000} 
\begin{equation}\label{eq:1}
    A_{\rm {ZF}}(t)=A_0e^{-({\lambda}t)^\beta},
\end{equation}
where \textit{$A_0$} is the initial asymmetry, $\lambda$ is the depolarization rate, $\beta$ is the stretching exponent, and \textit{t} is time. The parameter $\beta$ is roughly corresponding to  various magnetic states including paramagnetism ($\beta\approx2$), ordered/disordered magnetism ($\beta\approx1$) and spin-glass-like states or dynamically fluctuating spin systems ($\beta\approx0.5$)~\cite{Fowlie2022,Uemura1994,Cho2020}, see Fig.~\ref{fig:Figure2}(d).  The thicker superlattices with $m=2$ and $m=3$ exhibit expected behavior for an antiferromagnetic phase transition where at high temperatures, the values of $\beta$ are close to the paramagnetic value of $2$, and they decrease towards 1  with decreasing temperature as the static magnetic order sets in. In contrast, for the $m=1$  superlattice, $\beta$  is in the whole temperature range close to 0.5. The obtained values of $\beta$  depict the qualitative difference in the magnetic properties of the $m=1$ superlattice on the one hand and $m=2$ and $m=3$ superlattices on the other hand. A similar qualitative difference can be seen in the values of  $\lambda$, see Supplementary Fig.~\ref{fig:Figure4}(b).

\medskip

\begin{figure}[t]
    \centering
    \includegraphics[width=11cm]{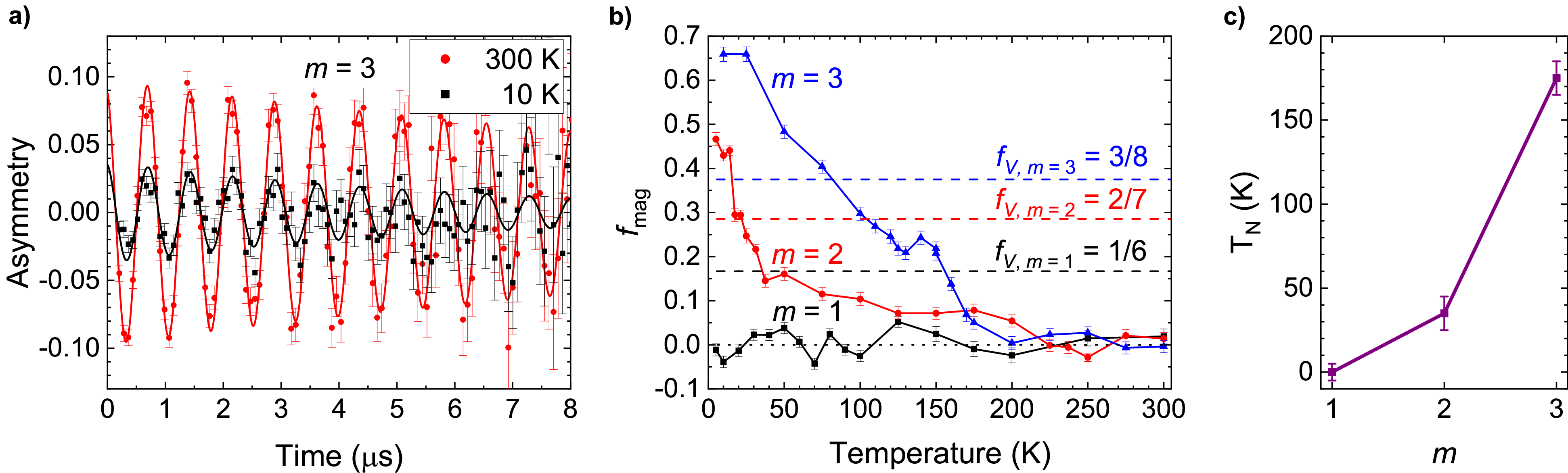}
    \caption{\textbf{The magnetic volume fraction and the Neel temperature}
    a)~Time evolution of the muon spin polarization in the weak transverse field  of 10~mT in the $m = 3$ superlattice  at 300~K and 10~K shown with a fit (solid lines) using Eq.~(\ref{eq:3}). (b)~Magnetic volume fraction, $f_{\rm mag}$, of [(LaFeO$_3$)$_m$/(SrTiO$_3$)$_5$]$_{10}$ superlattices evaluated from the measurement in the weak transverse field. Horizontal dashed  lines represent the volume fraction of LaFeO$_3$ in the superlattices. (c) The Neel temperature with respect to  $m$ determined from panel (b).}
    \label{fig:Figure3}
\end{figure}

\subsection*{The magnetic volume fraction and the Neel temperature}
Muon spin rotation spectroscopy offers a way to determine the volume fraction of a magnetically ordered phase using a measurement where a weak external field is applied transverse to the muon spins. In a paramagnetic state, the  fluctuation rate of  electronic moments is too high to influence the muon spin direction, and thus the muons precess due to the external magnetic field, which is observed as an oscillation of the asymmetry. Figure~\ref{fig:Figure3}(a) shows these  oscillations in the weak transverse field asymmetry of  the $m=3$ superlattice at 300~K, which is at this temperature in the paramagnetic state. The solid line represents a fit using the exponentially damped cosine function
\begin{equation}
    A_{\rm TF}(t)=A_{0}\,
    e^{-\lambda_{\rm TF} t}
    \cos{[\gamma_\mu B_{\rm ext}t+\phi]}\;,
\label{eq:3}
\end{equation}
where $A_{0}$ is the initial asymmetry, $\lambda_{\rm TF}$ is the depolarisation rate, $B_{\rm ext}$ is the applied transverse field, and $\phi$ relates to the initial muon spin polarisation.
In an  ordered magnetic phase, muon spins quickly depolarize because of the large static fields, which leads to the decrease of the oscillation amplitude, as can be seen  in the  asymmetry of the $m=3$ superlattice at 10~K, see Fig.~\ref{fig:Figure3}(a).
This reduction of the oscillation amplitude is a clear sign of the formation of a static magnetic order at low temperatures. The magnitude of this decrease  yields the  magnetic volume fraction, $f_{\rm mag}$, which was calculated as
\begin{equation}
    f_{\rm mag}(T)=
    1-\frac{A_{0}(T)}{A_{0}(T_{\rm high})}\;,
    \label{eq:3a}
\end{equation}
where $A_{0}(T_{\rm high})$ is the mean of the initial weak transverse field asymmetry above 250~K in the expected paramagnetic state.
We have determined $f_{\rm mag}$ of our superlattices using measurements in a transverse field of 10~mT applied in a perpendicular direction to the superlattice surface.  We corrected $f_{\rm mag}$ for the muonium formation in SrTiO$_3$; for details, see Supplementary Sec.~\ref{SOMwtf}.  

The obtained $f_{\rm mag}$ for the $m=3$ superlattice, see  Fig.~\ref{fig:Figure3}(b),   exhibits an onset near 175~K and increases with lowering the temperature, which is typical for a magnetically ordered state. At 10~K, $f_{\rm mag}$ is above 0.6, which is  more than the \lfo\ volume fraction, $f_{V,m=3}= 3/8$, which depicts that the antiferromagnetic state is well developed with some stray fields reaching into  \sto\ layers. The stray fields are likely caused by the small canting of  LaFeO$_3$ moments~\cite{Eibschutz1967}.
In the $m=2$ superlattice, $f_{\rm mag}(T)$ 
exhibits a weak increase below  200~K, a sharp onset below 35~K and reaches above 0.4 at 5~K. This value is again larger than \lfo\ volume fraction $f_{V,m=2}=2/7$, demonstrating that even in this superlattice with only two monolayers of \lfo, the antiferromagnetic  state is well developed at 5~K, although with  significantly reduced \tn\ to  35~K.  In contrast,  $f_{\rm mag}$  of the $m=1$ superlattice is zero within the experimental error bars down to the lowest measured temperature of 5~K, showing the absence of formation of a static order in the measured temperature range. 
The qualitative difference between $f_{\rm mag}$ of $m=3$ and $m=2$ superlattices on the one hand and of the $m=1$ superlattice on the other hand again depicts the qualitative difference in their magnetic ground state.

The dependence of \tn\ on $m$ is summarized in  Fig.~\ref{fig:Figure3}(c). Because  muons stop  in the superlattice at various sites, it is not possible to determine  from the  data whether the order is ferromagnetic or antiferromagnetic. However, we assume that the observed order is antiferromagnetic since its transition temperature increases with increasing $m$, and we expect that for large $m$,  the properties should approach those of  bulk \lfo. In our superlattices with $m\leq3$, \tn\ is still much smaller compared  to the bulk value of 740~K. To some extent, this reduction can be due to a change of valency of Fe due to proximity to Sr ions at the interface between \lfo\ and \sto. This effect is  the strongest in the $m=1$ superlattice where the iron oxide layer is formed only by one LaO and one FeO$_2$ layer, see Fig.~\ref{fig:Figure1}(a), and thus  Fe ions are surrounded equally by La and Sr ions. Nevertheless, since bulk La$_{0.5}$Sr$_{0.5}$FeO$_3$ is still antiferromagnetic with \tn\ of about 250~K \cite{Matsuno1999},  we conclude that the strong reduction of \tn\ of $m=2$ and $m=1$ superlattices is  predominantly due to the dimensional crossover rather than due to the change of the Fe valency. 

\begin{figure}[t]
    \centering
    \includegraphics[width=11cm]{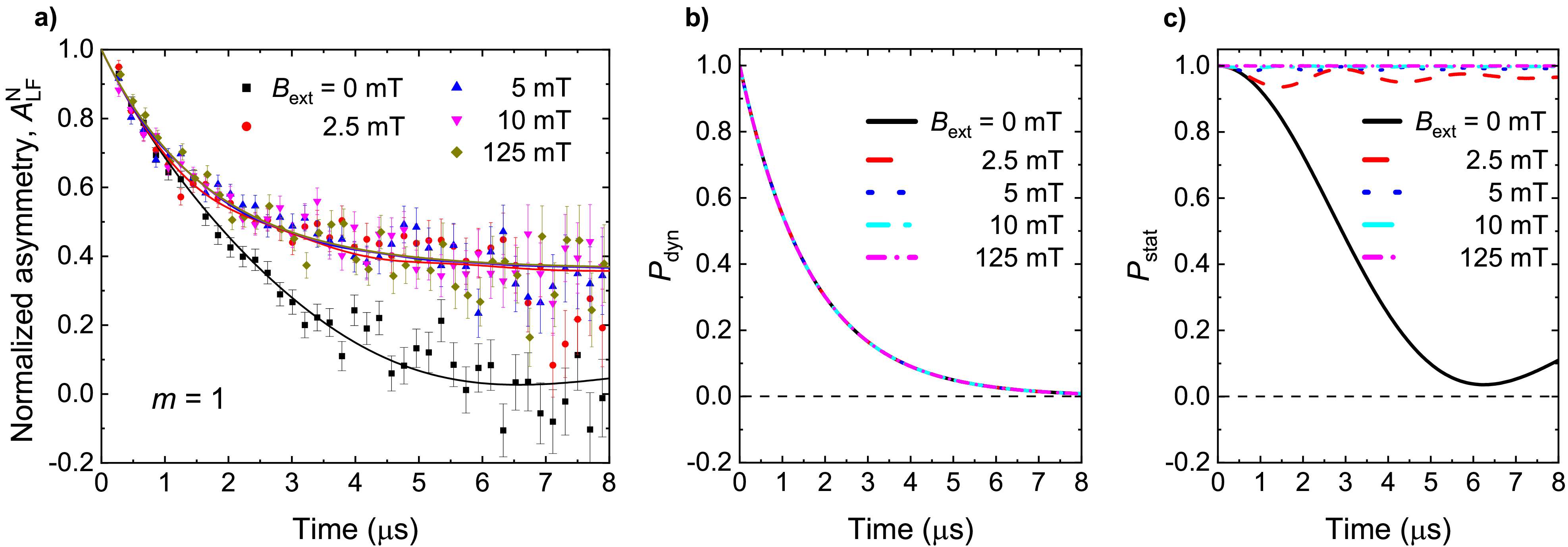}
    \caption{\textbf{Differentiation between the static and dynamic magnetism}
    (a)  Time evolution of  normalized muon spin polarization, $A^N_{\rm LF}$, of the $m=1$ superlattice at 5~K for various applied longitudinal magnetic fields. Error bars  represent one standard deviation. The solid lines represent fit using a model given by Eq.~(\ref{KT}). The significant decrease  of asymmetry at high fields is a hallmark of  dynamic magnetism. Panels (b) and (c) display the theoretical Gaussian Kubo-Toyabe functions used in the fit for dynamically fluctuating moments, $P_{\rm dyn}$, and for static disordered moments, $P_{\rm stat}$, respectively.  }
    \label{fig:Figure3b}
\end{figure}
\subsection*{Differentiation between the static and dynamic magnetism}

The zero field and the weak transverse field data indicate that there is no magnetic order in the $m=1$ superlattice down to 5~K. This could be explained by two scenarios: a static disorder (e.g., due to structural defects) or dynamic fluctuations of the electronic moments.  Muon spin rotation spectroscopy offers a way  to unequivocally differentiate between static magnetism and dynamically fluctuating fields by measurements in the  magnetic field longitudinal to the muon spin direction. In the presence of static magnetism, muons in the longitudinal field with a magnitude much larger than that of the local fields essentially do not precess (so-called decouple from the local fields) and thus do not depolarize  in contrast to the zero field measurements. However, if the local fields are fluctuating, they cause a random muon spin-flip (a transition between the Zeeman split energy levels) and cause the muon-spin depolarization even in the longitudinal field,  essentially the same as in zero field \cite{Blundell1999}. Time evolutions of muon spin polarization in the $m=1$ superlattice  at 5~K in several longitudinal fields are shown in Fig.~\ref{fig:Figure3b}(a); data are normalized as detailed in Supplementary Sec.~\ref{SOMlf}.  The asymmetry increases between  zero field and  2.5~mT, which is  caused by the decoupling of the muon spins from the static nuclear moments of \sto~\cite{Hayano1979}. However, for higher fields between 2.5~mT to 125~mT, the asymmetry is essentially field independent and exhibits at  8~$\mu$s considerable depolarization to about 40\% of the initial value. Such a significant depolarization independent of the longitudinal field  is a  hallmark of   fluctuating  electronic moments (see, e.g., Ref.~\cite{Balz2016}). 

We have modeled the normalized asymmetry in the longitudinal field, $A^N_{\rm LF}$,  as a sum of the theoretical Gaussian Kubo-Toyabe functions for dynamic fluctuations, $P_{\rm dyn}$ \cite{Keren1994} and for the static disorder, $P_{\rm stat}$ \cite{Hayano1979}
\begin{equation}
\label{KT}
    A^N_{\rm LF}=c\,P_{\rm dyn}+ 
    (1-c)\,P_{\rm stat}\;,
\end{equation}
where $c$ is the volume fraction of the fluctuating part;
for details see Supplementary Sec.~\ref{SOMlf}.
The global fit for all longitudinal fields $B_{\rm ext}$, see solid lines in  Fig.~\ref{fig:Figure3b}(a), yields the  volume fraction $c=0.64\pm 0.06$ and the distribution of the static disordered moments ${\sigma_{\rm s}}\slash{\gamma_\mu}=0.32\pm 0.08$~mT.
The functions $P_{\rm dyn}$ displayed in Fig.~\ref{fig:Figure3b}(b) for the obtained parameter values   are essentially field independent and vanish at 8~$\mu$s. In contrast, $P_{\rm stat}$, displayed in Fig.~\ref{fig:Figure3b}(c),  sensitively  depends on the external magnetic field. This difference allows the model to discern between static disorder and dynamically fluctuating moments. 
The obtained value of ${\sigma_{\rm s}}\slash{\gamma_\mu}=0.32\pm 0.08$~mT is typical for nuclear moments~\cite{Hayano1979}. It corresponds to regions in \sto\  where the dipolar fields from the iron moments are smaller compared to the nuclear fields. This area's volume fraction of $(1-c)=0.36$ corresponds to  the width of about 2.2 monolayers, presumably located in the middle of \sto\ layers. The fact that we can fit the data with the model yielding such a small value of ${\sigma_{\rm s}}\slash{\gamma_\mu}$ at all external fields  is incompatible with the picture of statically disordered iron moments with local fields expected to be in the order of 100-250~mT~\cite{Holzschuh1983}. If iron moments were static,  the increase of  the longitudinal field between 10 and 125~mT would lead to a significant increase in the asymmetry~\cite{Blundell1999}. The field-independent  asymmetry exhibiting such a considerable depolarization for fields above 2.5~mT can be explained only as a consequence of the fluctuating  iron moments.

In summary, the  muon spin rotation data in zero, transverse and longitudinal fields  consistently show that  (i) $m=3$ and $m=2$ superlattices exhibit a long-range  antiferromagnetic order with \tn\ of 175~K and 35~K, respectively,  (ii) that the magnetic properties of the $m=1$ superlattice are qualitatively different with no long-range order down to the lowest measured temperature of 5~K and (iii) that at this temperature, the electronic moments are fluctuating rather than statically disordered. These findings point towards a dimensional magnetic crossover where for the superlattice with a single monolayer of iron oxide, the static antiferromagnetic  order is lost due to enhanced magnitude of spin fluctuations, as expected from the  Mermin-Wagner theorem.

\backmatter






\subsection*{Methods}
\subsubsection*{Sample growth and characterization}
Superlattices were fabricated by pulsed laser deposition on $10\times10$~mm$^2$ TiO-terminated SrTiO$_3$ (001) substrates. The deposition temperature of the substrates was 570~$^\circ$C, and the background oxygen pressure was 0.01~mbar. The thickness of layers  was \textit{in situ} controlled by reflection of high-energy electron diffraction. The samples were annealed \textit{ex situ}  in an oxygen atmosphere at 550~$^\circ$C to reduce the concentration of oxygen vacancies. We fabricated sets of 3-4 samples of each superlattice that formed a sample mosaic to improve the signal-to-noise ratio of the muon spin rotation data. The structural quality of the superlattices was characterized using an atomic force microscope (Bruker Dimension Icon) and X-ray diffractometer (Rigaku Smartlab). Atomic force microscope images were analyzed by Gwyddion software \cite{Necas2012} and the superlattice structure shown in Fig. \ref{fig:Figure1}(a) by VESTA software \cite{Momma2011}.

\subsubsection*{Low-energy muon spin rotation}
Low-energy muon spin rotation experiments were  performed at the $\mu$E4 beamline of the Swiss Muon Source at Paul Scherrer Institute, Villigen. We have used 2~keV muon  beam   that results in  an implantation profile, where most of the muons stop in the superlattices, see Supplementary Fig.~\ref{fig:Figure9}(b). $\mu$SR data were analyzed using musrfit \cite{Suter2012}.
\subsection*{Acknowledgements}
We thank C. Bernhard for the fruitful discussions. We thank K. Bernatova (Tescan Orsay Holding) and S. Dinara for their help with sample preparation. A.D. and M.K. acknowledge  the financial support by the MEYS of the Czech Republic under the
project CEITEC 2020 (LQ1601), by the Czech Science Foundation (GACR) under Project No. GA20-10377S and CzechNanoLab project LM2018110 funded by MEYS CR for the financial support of the measurements/sample fabrication at CEITEC Nano Research Infrastructure.
\end{bibunit}
\begin{bibunit}
\newpage

\begin{center}
    \Large
    \textbf{Supplementary online material}
\end{center}
\setcounter{figure}{0}
\renewcommand{\figurename}{Supplementary Fig.}
\setcounter{section}{0}

\section{Sample characterization }\label{SOMsample}
The X-ray diffraction data shown in Fig.~\ref{fig:Figure1}(c) exhibit the superlattice  diffraction peaks (SL$_{-1}$ and SL$_{1}$) of the (\lfo)$_m$+(\sto)$_5$ bilayer. The thickness of the bilayer was determined from the angular position of these diffraction peaks using the Bragg equation. Supplementary Fig.~\ref{fig:Figure9}(a)  shows the obtained bilayer thickness as a function of  $m$, which is in good agreement with the thickness expected from the bulk lattice constants of SrTiO$_3$, $a_{\text{SrTiO}_3}=3.905$~\AA, and  out-of-plane pseudo-cubic lattice constant of epitaxial LaFeO$_3$ film, $a_{\text{LaFeO}_3}=4.009$~\AA. 

\section{Low energy muon spin rotation }\label{SOMlem}

The low-energy muon spin rotation (LE$\mu$SR) spectroscopy~\cite{Prokscha2008,Suter2023}  with relatively small muon penetration depth compared to the standard muon spin rotation ($\mu$SR)  enables studying  thin films and heterostructures. We have used the  muon beam with  2~keV that provides an implantation profile where most of the muons stop in the superlattices; see Supplementary Fig.~\ref{fig:Figure9}(b). The implantation profile was  calculated using  Monte Carlo  TRIM.SP code \cite{Eckstein1991}. Data in all measurements are analyzed above 0.1~$\mu$s since below, strong systematic error due to back reflection of muons arise~\cite{Suter2023}.

\begin{figure}[b]
    \centering
    \includegraphics[width=11.5cm]{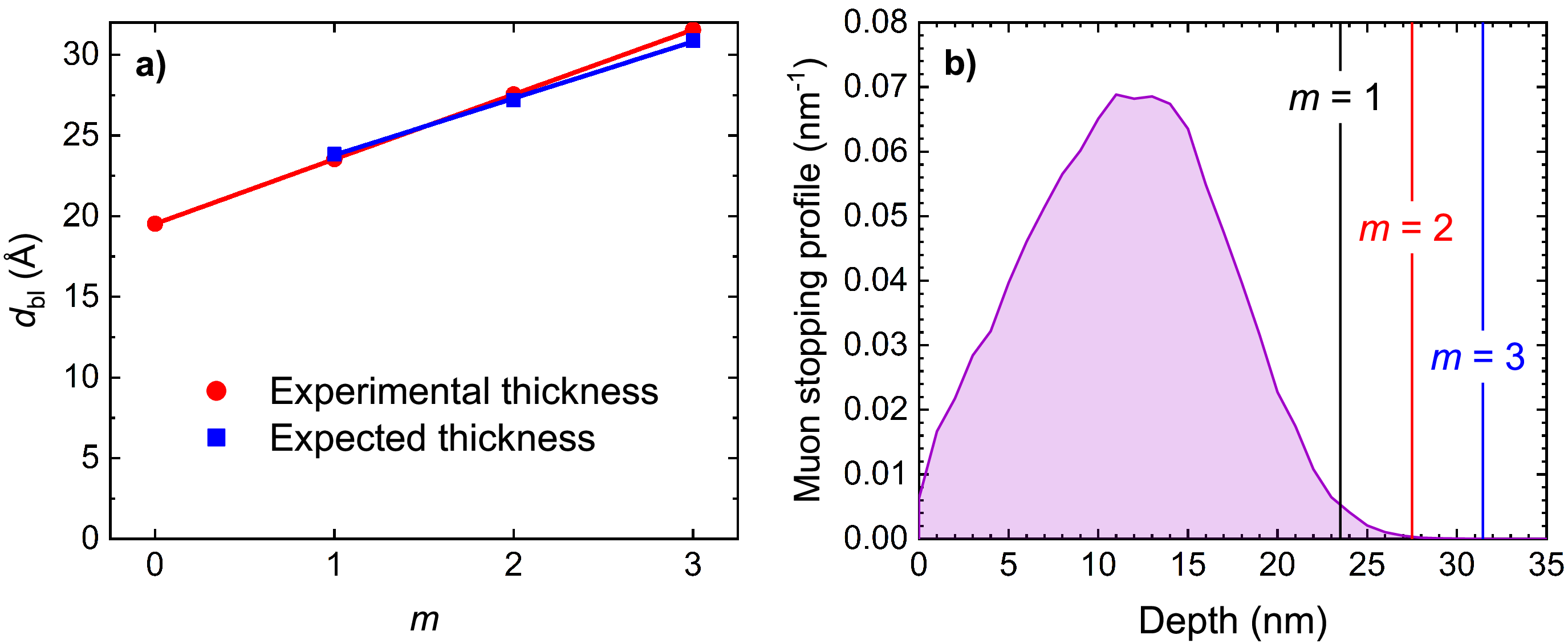}
    \caption{(a) Comparison of the  thickness of (\lfo)$_m$+(\sto)$_5$ bilayer, 
    $d_{\rm bl}$, determined from X-ray diffraction and the one expected from SrTiO$_3$ and  LaFeO$_3$ lattice constants as detailed in the text. (b) Simulated implantation profile for a muon beam energy of 2~keV. Vertical lines mark the thicknesses of [(LaFeO$_3$)$_m$/(SrTiO$_3$)$_5$]$_{10}$ superlattices. }
    \label{fig:Figure9}
\end{figure}

\subsection{Zero field muon spin rotation }\label{SOMzf}
The zero-field asymmetry spectra presented in Fig.~\ref{fig:Figure2}(a)-(c) do not exhibit any oscillatory time evolution typically seen in bulk crystals. This arises because  the structure of our superlattices leads to a broad distribution of internal fields, including the stray fields due to the iron spin canting that spread through  SrTiO$_3$ layers. This  corresponds to a large distribution of Larmor frequencies inevitably leading to a fast damping of the oscillations.

\begin{figure}[t]
    \centering
    \includegraphics[width=11cm]{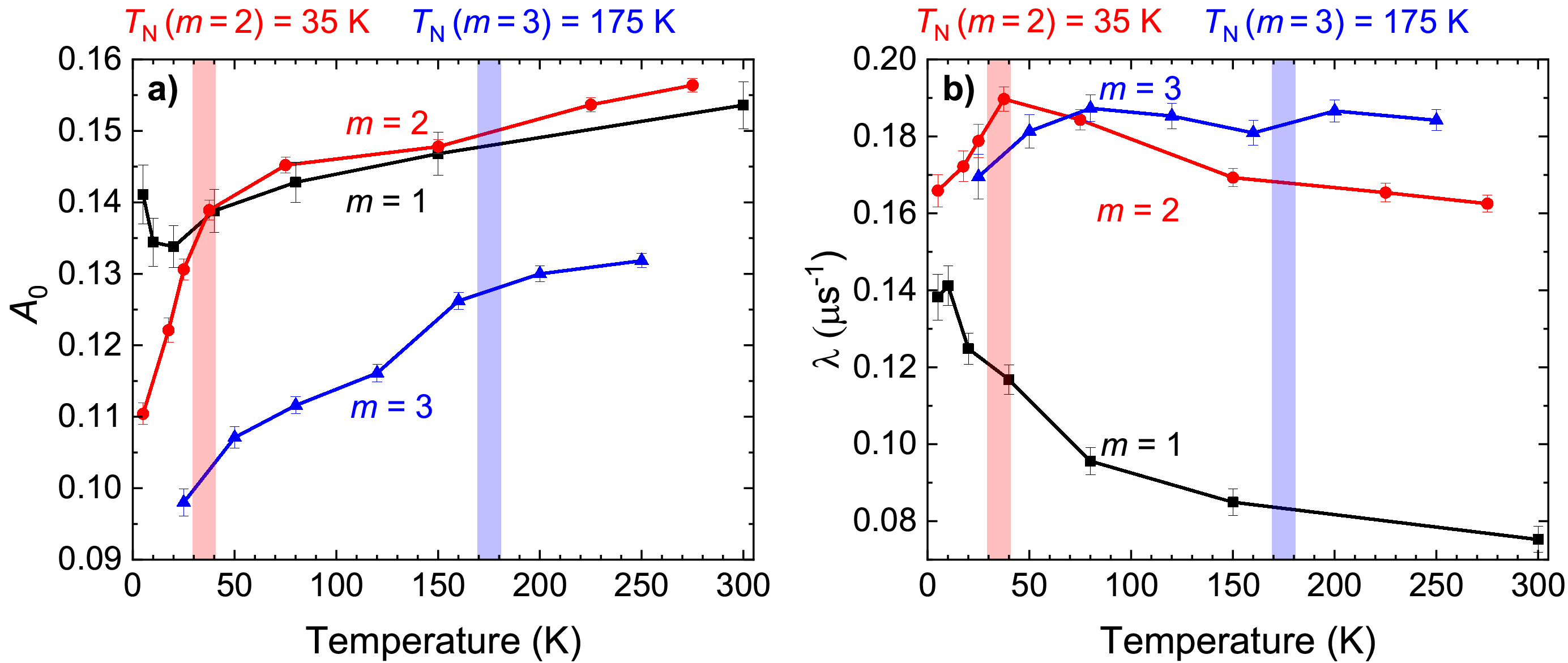}
    \caption{ (a) Initial asymmetry,  $A_0$, and (b) depolarization rate, $\lambda$,  obtained from the fit of zero field data using Eq.~(\ref{eq:1}). The highlighted areas mark  Neel temperature of $m=2$ and $m=3$ superlattices.}
    \label{fig:Figure4}
\end{figure}

The time evolution of the zero-field asymmetry  presented in Fig.~\ref{fig:Figure2}(a)-(c) was fitted using the stretched exponential function~(\ref{eq:1}) whose exponent $\beta$ is shown in Fig.~\ref{fig:Figure2}(d). For completeness, values of the other fitted parameters, i.e., the  initial asymmetry, $A_0$, and the depolarization rate, $\lambda$,  are shown in Supplementary Figs.~\ref{fig:Figure4}(a) and ~\ref{fig:Figure4}(b), respectively. The temperature dependence of $A_0$ exhibits a noticeable decrease  below  \tn, that is, below 175~K for the $m=3$ and below 35~K for the $m=2$ superlattice. This decrease is expected in an ordered magnetic phase where the muons quickly depolarize due to strong static local fields. In  the $m=1$ superlattice,  $A_0$ exhibits only a gradual and relatively weak decrease with decreasing temperature without a sharper onset in agreement with the interpretation that there is no static order in this superlattice. Surprisingly, $A_0$ of the $m=1$  superlattice seems to increase from 10 to 5~K. However, this increase is on the level of one standard deviation, and we do not consider it significant enough. 

The temperature dependence of $\lambda$,  shown in Supplementary Fig.~\ref{fig:Figure4}(b), is roughly constant for  $m = 3$  and $m = 2$ superlattices with values of $\lambda$ in the range from 0.16 to 0.19 $\mu$s$^{-1}$. It exhibits only a small indication of a decrease with decreasing temperature below \tn, particularly in the $m = 2$ superlattice, connected with the formation of the static magnetic order. In contrast, in the  $m = 1$ superlattice, the values of  $\lambda$ are  above 100~K  about two times smaller compared to  $m = 3$  and $m = 2$ superlattices. In addition,  $\lambda$ significantly increases with decreasing temperature below 100~K to about 0.14~$\mu$s$^{-1}$ at 5~K compared to  0.075~$\mu$s$^{-1}$ at 300~K. This almost doubling of the depolarization rate with decreasing temperature is a strong indication of a decrease in the electronic fluctuation rate. The latter is a typical signature of a fluctuating magnetic ground state \cite{Clark2013}. The qualitative and quantitative difference of the temperature dependence of  $\lambda$ between $m = 3$  and $m = 2$ superlattices on the one hand and the $m = 1$ superlattice on the other hand  again depicts the difference in their respective magnetic ground states. Note that in the case of the $m=2$ superlattice, the temperature dependence of $\lambda$ exhibits a clear maximum at \tn\ of about 35~K with a significant decrease below. This behavior is expected for a static magnetic order that sets in  at a phase transition temperature \cite{Grutter2018}. 

\begin{figure}[t]
    \centering
    \includegraphics[width=7cm]{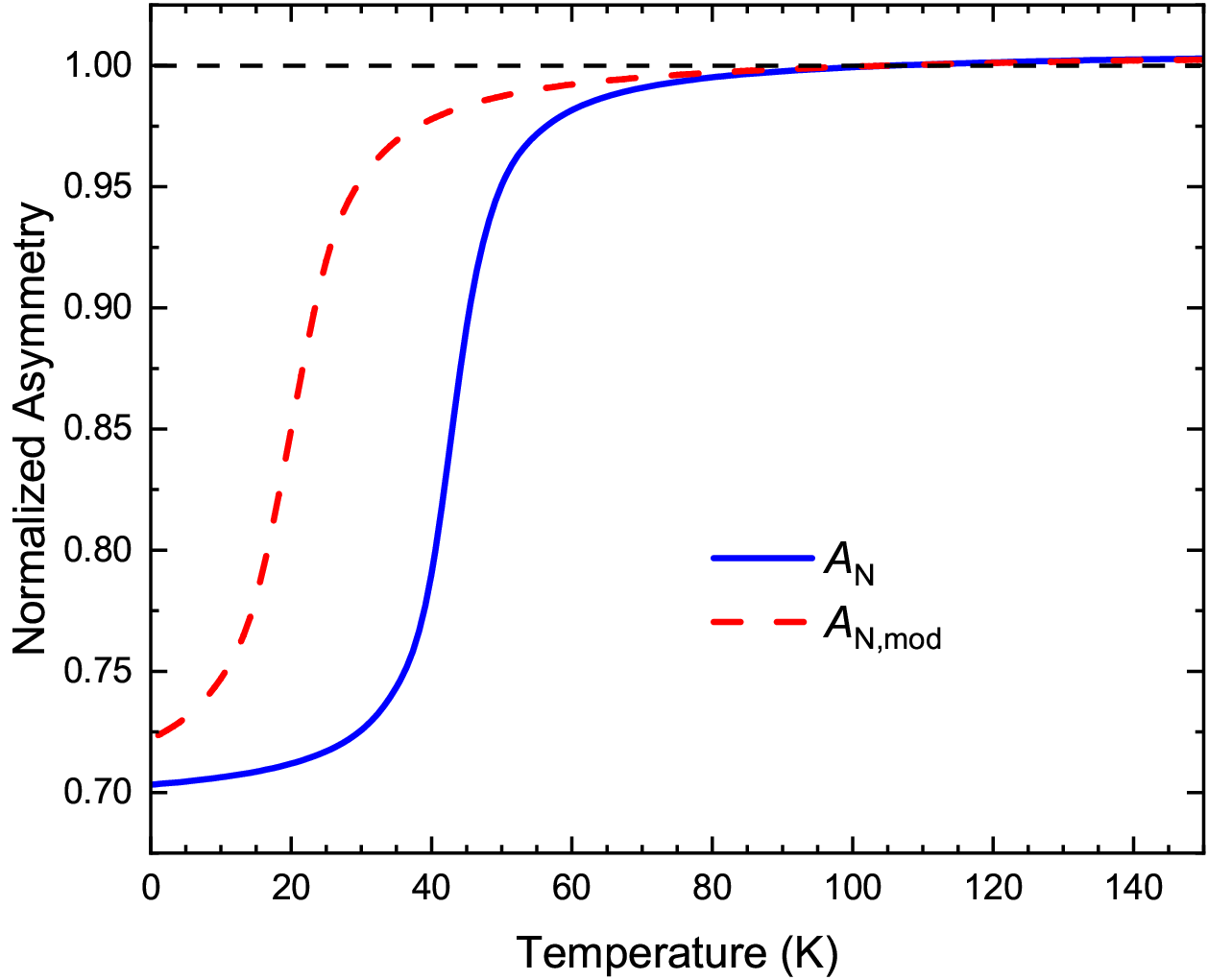}
    \caption{Normalized asymmetry due to the muonium formation in \sto\ expressed by Eq.~(\ref{eq:4}) (blue solid line) and Eq.~(\ref{eq:6}) (red dashed line).}
    \label{fig:Figure5}
\end{figure}

\subsection{Weak transverse field muon spin rotation }\label{SOMwtf}
 In the weak transverse field data analysis, we shall consider  the formation   of muonium (a bound state of a muon and an electron) in \sto. Because the muon spin in muonium precesses at a different frequency than a free muon spin, the formation of muonium occurring  below about 50~K in \sto~\cite{Salman2014}  influences the weak transverse field data of our superlattices. The temperature dependence of the normalized weak transverse field asymmetry  for \sto\ at 1.6~keV implanting energy was described by the empiric equation~\cite{Salman2014} 
\begin{equation}
    A_{\rm N}(T)=0.1\arctan{\frac{T-43}{4.427}+0.85}\;,
\label{eq:4}
\end{equation}
see Supplementary Fig.~\ref{fig:Figure5}.
Our measurements were performed at the implanting muon energy of 2~keV, which is close  enough to use Eq.~(\ref{eq:4}) as a starting point in muonium correction.
Assuming that the muonium is formed only in \sto\ layers, the depolarization due to the muonium formation is subtracted from the data using the following equation  
\begin{equation}
\label{eq:5}
    A_{0,c}(T)=A_0(T)+
    f_{\rm SrTiO_3}\left[1-A_{\rm N}(T)
    \right]A_0(T_{\rm high})\;,
\end{equation}
 where $f_{\rm SrTiO_3}$ is the volume fraction of \sto\ in a given superlattice, and $A_0(T_{\rm high})$ is the asymmetry at high enough temperature. For $A_0(T_{\rm high})$ we have used a mean value above 250~K where the superlattices are  in the paramagnetic state and  the influence of muonium is negligible.
The correction for the muonium formation is significant only below the temperature of the muonium formation of about 50~K,  where   $A_{\rm N}(T)$ is significantly smaller than unity, see Supplementary Fig.~\ref{fig:Figure5}. At higher temperatures, $A_{\rm N}(T)\approx1$ and  the second term on the right-hand side of  Eq.~(\ref{eq:5}) vanishes.

\begin{figure}[t]
    \centering
    \includegraphics[width=12cm]{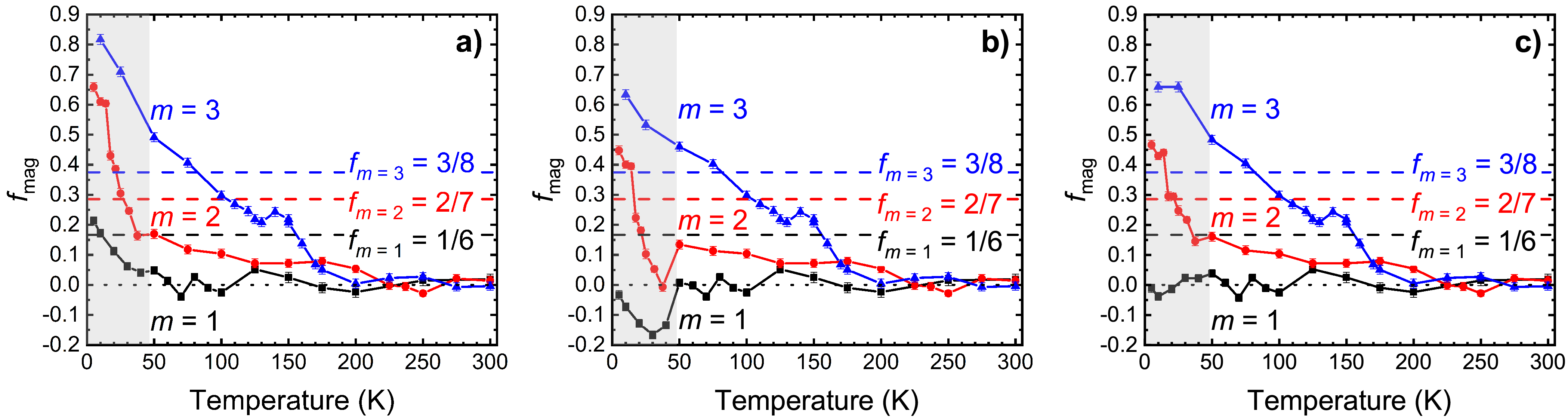}
    \caption{Magnetic volume fractions obtained from the weak transverse field measurement. Panel (a) shows  values   obtained  without the correction for  muonium formation,  and  panels (b) and (c) show those  corrected for the muonium formation  using Eq.~(\ref{eq:4}) and  Eq.~(\ref{eq:6}), respectively.}
    \label{fig:Figure6}
\end{figure}

The  magnetic volume fraction of the superlattice, $f_{\rm mag}(T)$, is calculated as~\cite{Fowlie2022}
\begin{equation}
    f_{\rm mag}(T)=
    1-\frac{A_{0,c}(T)}{A_{0,c}(T_{\rm high})}\;,
    \label{eq:6}
\end{equation}
 where $A_{0,c}(T_{\rm high})$ is the mean of the initial weak transverse field asymmetry above 250~K in the expected paramagnetic state. Supplementary Fig.~\ref{fig:Figure6}(a) shows  $f_{\rm mag}$ calculated without the muonium correction (using $A_{\rm N}(T)=1$). The shaded regions show temperatures below about 50~K where the muonium formation takes place.
Values of $f_{\rm mag}$   corrected for the muonium formation using  Eq.~(\ref{eq:4}), see Supplementary Fig.~\ref{fig:Figure6}(b),  suddenly decrease   below about 50~K for superlattices with $m = 2$ and $m = 1$, which leads, for the case of $m=1$, even to  nonphysical values significantly below zero. Most likely, the step-like correction for muonium formation using Eq.~(\ref{eq:4}), which was obtained on \sto\ single crystal, is sharper and centered at a different temperature than what would be appropriate for ultrathin \sto\ layers of our  superlattices. We have therefore adjusted the temperature and width of the transition in Eq.~(\ref{eq:4}) where the muonium formation occurs so that  $f_{\rm mag}$ is not negative for the superlattice $m = 1$. This approach yielded 
\begin{equation}
    A_{N, \rm mod}(T)=0.1\arctan{\frac{T-20}{6}+0.85}\;,
\label{eq:7}
\end{equation}
see Supplementary Fig.~\ref{fig:Figure5}(b). Corresponding $f_{\rm mag}$ is shown  in Supplementary Fig.~\ref{fig:Figure6}(c) and in the main part of the paper in Fig.~\ref{fig:Figure3}(b). Note that we did not adjust the multiplication factor of the step-like arctan function in Eq.~(\ref{eq:7})  corresponding to the magnitude of the correction. Consequently, the values of $f_{\rm mag}$ at 5~K resulting from  the two corrections [cf. Supplementary Figs.~\ref{fig:Figure6}(b) and~\ref{fig:Figure6}(c)] are almost the same. Similarly, the  main  conclusions are the same: the magnetic volume fraction in the $m=1$ superlattice  at 5~K is essentially zero corresponding to the absence of a static order formed in the measured temperature range in contrast to  the $m=2$ superlattice where it is significantly above zero (above 0.4 and above \lfo\ volume fraction of 2/7) and thus the superlattice exhibits a static antiferromagnetic order.

\def\bext{\ensuremath{B_{\rm ext}}}
\subsection{Longitudinal field muon spin rotation}\label{SOMlf}
\begin{figure}[h]
    \centering
    \includegraphics[width=7cm]{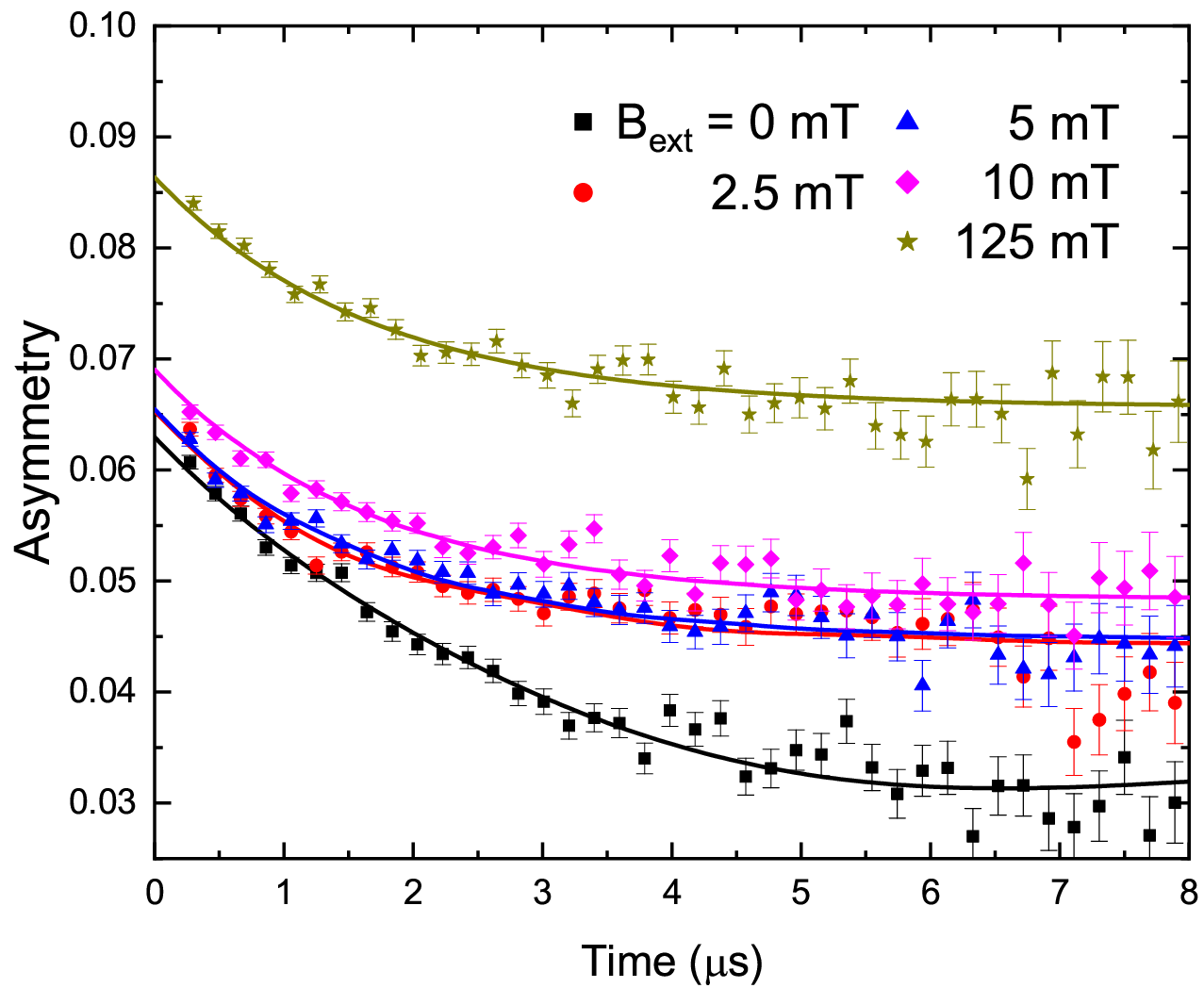}
    \caption{Time evolution of the asymmetry from longitudinal field measurement of the $m=1$ superlattice at 5~K  for several applied magnetic fields displayed together with a model (solid lines), see Eq.~(\ref{eq:8}). The error bars denote one standard deviation.}
    \label{fig:Figure7}
\end{figure}

Supplementary Fig.~\ref{fig:Figure7} shows the asymmetry of the $m=1$ superlattice measured at 5~K for several  longitudinal fields. The data are analyzed with the model
\begin{equation}
    A_{\rm LF}(t, \bext)=A\left[ c\,P_{\rm dyn}(t,\bext)+
    (1-c)P_{\rm stat}(t,\bext)
    \right]+A_0(\bext)\;.
\label{eq:8}
\end{equation}
Here $A_0(\bext)$  is a background asymmetry which, in principle, depends on magnetic field \bext. In LE${\mu}$SR, muons are focused onto the sample by the external magnetic field,  and thus different magnetic fields give rise to a different background. $A$ is the normalization constant that is field independent. The depolarization due to the sample is modeled as a weighted average of the theoretical Gaussian Kubo-Toyabe functions for the static disorder,  $P_{\rm stat}$~\cite{Hayano1979},

\begin{equation}
    P_{\rm stat}(B_{\rm ext}=0, t)=
    \frac{1}{3}+ \frac{2}{3}
    (1-\sigma_{\rm s}^2t^2)\exp{\left[-\frac{\sigma_{\rm s}^2t^2}{2} \right]\;,}
\label{eq:9}
\end{equation}

\begin{equation}
\begin{split}
    P_{\rm stat}(B_{\rm ext}, t)=
    \,&1-\frac{2\sigma_{\rm s}^2}{(\gamma_\mu B_{\rm ext})^2}\left[1-\exp{\left(-\frac{\sigma_{\rm s}^2 t^2}{2}\right)}\cos{\left(\gamma_{\mu} B_{\rm ext}t\right)}\right]+\\
    &+\frac{2\sigma_{\rm s}^4}{(\gamma_\mu B_{\rm ext})^3}\int_{0}^t\exp{\left(\frac{\sigma_{\rm s}^2 \tau^2}{2}\right)}\sin{(\gamma_\mu B_{\rm ext}\tau)}d\tau\;,
\label{eq:10}
\end{split}
\end{equation}

and the dynamic fluctuation, $P_{\rm dyn}$~\cite{Keren1994}, 

\begin{equation}
    P_{\rm dyn}(B_{\rm ext}, t)=
    \exp{\left[-\frac{2\sigma_{\rm d}^2\nu}{({\gamma_{\mu}}B_{\rm ext})^2+\nu^2}t\right]}\;,
\label{eq:11}
\end{equation}
where $\nu$ is the fluctuation rate. 
The depolarization rate $\sigma$ appearing in Eqs.~(\ref{eq:9})-(\ref{eq:11}) is defined as 
$\sigma=\gamma_\mu\sqrt{<{\Delta B}^2>}$ where $<{\Delta B}^2>$ is the second moment of the field distribution. The subscripts `$\rm s$' and `$\rm d$' denote whether the second moment corresponds to  the static or dynamic magnetic field distribution.  The volume fraction of  the dynamically fluctuating  part is expressed by the field-independent parameter $c$.

\begin{table}
\centering
\begin{tabular}{l l }
 \hline\hline
 Variable & Value \\
 \hline
 $ {\sigma_{\rm s}}\slash{\gamma_\mu}$ [mT] & 0.32 $\pm$ 0.08  \\
 ${\sigma_{\rm d}}\slash{\gamma_\mu}$ [mT] fixed & 250  \\
 $\nu$ [GHz] & 150 $\pm$ 20  \\
 $c$  & 0.64 $\pm$ 0.06  \\
 $A$  & 0.032 $\pm$ 0.003  \\
 $A_0(B_{\rm ext}=0$~mT)  & 0.030 $\pm$ 0.002  \\
 $A_0(B_{\rm ext}=2.5$~mT)  & 0.032 $\pm$ 0.003  \\
  $A_0(B_{\rm ext}=5$~mT)  & 0.032 $\pm$ 0.003  \\
$A_0(B_{\rm ext}=10$~mT)  & 0.036 $\pm$ 0.003  \\
 $A_0(B_{\rm ext}=125$~mT)  & 0.053 $\pm$ 0.003  \\
\hline\hline
\end{tabular}
\caption{Values of parameters obtained from the global fit of the longitudinal field data by Eq.~(\ref{eq:8}). The errors represent one standard deviation.}
\label{table:1}
\end{table}

The data  shown in Supplementary Fig.~\ref{fig:Figure7} (points) were analyzed with a global fit of asymmetries at all measured fields with the model expressed by Eq.~(\ref{eq:8}) (solid lines). The obtained values of parameters are shown in Tab.~\ref{table:1}. Because of the high correlation between $\sigma_{\rm dyn}$ and the fluctuation rate $\nu$, we fixed the distribution width of the  dynamically fluctuating moments,  ${{\sigma_{\rm dyn}}\slash{\gamma_\mu}}$,  to  250~mT since similar values  of the internal fields were observed in orthoferrites~\cite{Holzschuh1983}.  The corresponding  value of the fluctuation rate is  $\nu= 150 \pm 20 $~GHz. The obtained value ${\sigma_{\rm s}}\slash{\gamma_\mu}=0.32\pm 0.08$~mT exhibits  a relatively large error that is due to the comparably large field of 2.5~mT used in the measurements. To determine  this value with better precision, one would need to measure with significantly smaller fields. The errors  of other  values shown in Tab.~\ref{table:1}  are reasonably low (about 10\%), which demonstrates that the global fit is well conditioned. Particularly, the fit allowed us to  determine the constants $A$ and $A_0(\bext)$ with reasonable precision. For the sake of simplicity, we display  in Fig.~\ref{fig:Figure3b}  the data as normalized asymmetry 
\begin{equation}
    A_{\rm LF}^N(t, \bext)=[A_{\rm LF}(t, \bext) - A_0(\bext)]/A\;.
\label{equation:12}
\end{equation}

\newpage

\section{Run logs}\label{sec5}

\begin{table}[!h]
    \centering
    \begin{tabular}{|c| c |c |c |c |c|}
    \hline
    Measurement & \textit{T} [K] & \textit{E} [keV] & \textit{B} [mT] & run no. & year \\[0.5ex]
    \hline
    \multirow{7}{3em}{Zero field} & 300 & 1.96 & 0 & 4664 & 2022\\
    & 150 & 1.96 & 0 & 4693 & 2022\\
    & 80 & 1.96 & 0 & 4665 & 2022\\
    & 40 & 1.96 & 0 & 4666 & 2022\\
    & 20 & 1.96 & 0 & 4667 & 2022\\
    & 10 & 1.96 & 0 & 4668 & 2022\\
    & 5 & 1.96 & 0 & 4669 & 2022\\[0.5ex]
    \hline
    \multirow{16}{3em}{Weak transverse field} & 300 & 2.01 & 10 & 6608 & 2021\\
    & 250 & 2.01 & 10 & 6609 & 2021\\
    & 200 & 2.01 & 10 & 6610 & 2021\\
    & 175 & 2.01 & 10 & 6611 & 2021\\
    & 150 & 2.01 & 10 & 6612 & 2021\\
    & 125 & 2.01 & 10 & 6613, 6607 & 2021\\
    & 100 & 2.01 & 10 & 6614, 6596 & 2021\\
    & 90 & 2.01 & 10 & 6615, 6597 & 2021\\
    & 80 & 2.01 & 10 & 6616, 6598 & 2021\\
    & 70 & 2.01 & 10 & 6617, 6599 & 2021\\
    & 60 & 2.01 & 10 & 6618, 6600 & 2021\\
    & 50 & 2.01 & 10 & 6619, 6601 & 2021\\
    & 40 & 2.01 & 10 & 6620, 6602 & 2021\\
    & 30 & 2.01 & 10 & 6621, 6603 & 2021\\
    & 20 & 2.01 & 10 & 6622, 6604 & 2021\\
    & 10 & 2.01 & 10 & 6623, 6605 & 2021\\
    & 5 & 2.01 & 10 & 6624, 6606 & 2021\\[0.5ex]
    \hline
    \multirow{5}{5.5em}{Longitudinal field} & 5 & 1.96 & 0 & 4677, 4678 & 2022\\
    & 5 & 1.96 & 2.5 & 4679, 4680 & 2022\\
    & 5 & 1.96 & 5 & 4681, 4682 & 2022\\
    & 5 & 1.96 & 10 & 4683, 4684 & 2022\\
    & 5 & 1.96 & 125 & 4695, 4696 & 2022\\[0.5ex]
    \hline
    \end{tabular}
    \label{table:2}
    \caption{Low energy $\mu$SR run log  for the $m = 1$ superlattice measured in  zero, weak transverse, and longitudinal fields.}
\end{table}

\begin{table}[!h]
\centering
\begin{tabular}{|c| c |c |c |c |c|}
 \hline
 Measurement & \textit{T} [K] & \textit{E} [keV] & \textit{B} [mT] & run no. & year \\[0.5ex]
 \hline
 \multirow{8}{4em}{Zero field} & 275 & 2.01 & 0 & 4833 & 2021\\
 & 225 & 2.01 & 0 & 4832 & 2021\\
 & 150 & 2.01 & 0 & 4831 & 2021\\
 & 75 & 2.01 & 0 & 4830 & 2021\\
 & 37.5 & 2.01 & 0 & 4829 & 2021\\
 & 25 & 2.01 & 0 & 4828 & 2021\\
 & 17.5 & 2.01 & 0 & 4826 & 2021\\
 & 5 & 2.01 & 0 & 4827 & 2021\\[0.5ex]
 \hline
 \multirow{20}{4em}{Weak transverse field} & 300 & 2.01 & 10 & 4823 & 2021\\
 & 275 & 2.01 & 10 & 4822 & 2021\\
 & 250 & 2.01 & 10 & 4834 & 2021\\
 & 237 & 2.01 & 10 & 4821 & 2021\\
 & 225 & 2.01 & 10 & 4820 & 2021\\
 & 200 & 2.01 & 10 & 4819 & 2021\\
 & 175 & 2.01 & 10 & 4818 & 2021\\
 & 150 & 2.01 & 10 & 4817 & 2021\\
 & 125 & 2.01 & 10 & 4816 & 2021\\
 & 100 & 2.01 & 10 & 4815 & 2021\\
 & 75 & 2.01 & 10 & 4814 & 2021\\
 & 50 & 2.01 & 10 & 4813 & 2021\\
 & 37.5 & 2.01 & 10 & 4824 & 2021\\
 & 31.27 & 2.01 & 10 & 4835 & 2021\\
 & 25 & 2.01 & 10 & 4812 & 2021\\
 & 21.3 & 2.01 & 10 & 4836 & 2021\\
 & 17.5 & 2.01 & 10 & 4825 & 2021\\
 & 14 & 2.01 & 10 & 4837 & 2021\\
 & 10 & 2.01 & 10 & 4810 & 2021\\
 & 5 & 2.01 & 10 & 4811 & 2021\\[0.5ex]
 \hline
\end{tabular}
\caption{Low energy $\mu$SR run log for the $m = 2$ superlattice   measured in zero and weak transverse fields.}
\label{table:3}
\end{table}

\begin{table}[!h]
\centering
    \begin{tabular}{|c| c |c |c |c |c|}
        \hline
         Measurement & \textit{T} [K] & \textit{E} [keV] & \textit{B} [mT] & run no. & year \\[0.5ex]
         \hline
         \multirow{7}{4em}{Zero field} & 250 & 1.96 & 0 & 4158 & 2021\\
         & 200 & 1.96 & 0 & 4155 & 2021\\
         & 160 & 1.95 & 0 & 4156 & 2021\\
         & 120 & 1.95 & 0 & 4157 & 2021\\
         & 80 & 1.95 & 0 & 4152 & 2021\\
         & 50 & 1.96 & 0 & 4154 & 2021\\
         & 25 & 1.95 & 0 & 4153 & 2021\\[0.5ex]
         \hline
        \multirow{20}{4em}{Weak transverse field} & 320 & 2.01 & 10 & 4140 & 2021\\
        & 300 & 2.01 & 10 & 4139 & 2021\\
        & 275 & 2.01 & 10 & 4138 & 2021\\
        & 250 & 2.01 & 10 & 4137 & 2021\\
        & 225 & 2.01 & 10 & 4136 & 2021\\
        & 200 & 2.01 & 10 & 4135 & 2021\\
        & 175 & 2.01 & 10 & 4134 & 2021\\
        & 170 & 2.01 & 10 & 4141 & 2021\\
        & 160 & 2.01 & 10 & 4142 & 2021\\
        & 150 & 2.01 & 10 & 4143,4133 & 2021\\
        & 140 & 2.01 & 10 & 4144 & 2021\\
        & 130 & 2.01 & 10 & 4145 & 2021\\
        & 125 & 2.01 & 10 & 4132 & 2021\\
        & 120 & 2.01 & 10 & 4146 & 2021\\
        & 110 & 2.01 & 10 & 4147 & 2021\\
        & 100 & 2.01 & 10 & 4131 & 2021\\
        & 75 & 2.01 & 10 & 4148 & 2021\\
        & 50 & 2.01 & 10 & 4149 & 2021\\
        & 25 & 2.01 & 10 & 4150 & 2021\\
        & 10 & 2.01 & 10 & 4151 & 2021\\[0.5ex]
        \hline
    \end{tabular}
\caption{Low energy $\mu$SR run log for the $m = 3$ superlattice  measured in zero and weak transverse fields.}
\label{table:4}
\end{table}

\end{bibunit}
\end{document}